\documentclass{article}

\usepackage{microtype}
\usepackage{graphicx}
\usepackage{subfigure}
\usepackage{booktabs} 
\usepackage{multirow}
\usepackage{multicol}
\usepackage{makecell}
\usepackage{amsmath}
\usepackage{amssymb}
\pdfoutput=1

\usepackage{hyperref}

\usepackage[accepted]{icml2021}

\icmltitlerunning{SoundDet: Polyphonic Moving Sound Event Detection and Localization from Raw Waveform}

\begin{document}

\twocolumn[
\icmltitle{SoundDet: Polyphonic Moving Sound Event Detection and Localization from Raw Waveform}




\icmlsetsymbol{equal}{*}

\begin{icmlauthorlist}
\icmlauthor{Yuhang He}{to}
\icmlauthor{Niki Trigoni}{to}
\icmlauthor{Andrew Markham}{to}
\end{icmlauthorlist}

\icmlaffiliation{to}{Department of Computer Science, University of Oxford, Oxford, United Kingdom. Email: firstname.lastname@cs.ox.ac.uk}

\icmlcorrespondingauthor{Andrew Markham}{andrew.markham@cs.ox.ac.uk}

\icmlkeywords{Machine Learning, ICML}

\vskip 0.3in
]



\printAffiliationsAndNotice{}  

\begin{abstract}
We present a new framework SoundDet, which is an end-to-end trainable and light-weight framework, for polyphonic moving sound event detection and localization. Prior methods typically approach this problem by preprocessing raw waveform into time-frequency representations, which is more amenable to process with well-established image processing pipelines. Prior methods also detect in segment-wise manner, leading to incomplete and partial detections. SoundDet takes a novel approach and directly consumes the raw, multichannel waveform and treats the spatio-temporal sound event as a complete ``sound-object" to be detected. Specifically, SoundDet consists of a backbone neural network and two parallel heads for temporal detection and spatial localization, respectively. Given the large sampling rate of raw waveform, the backbone network first learns a set of phase-sensitive and frequency-selective bank of filters to explicitly retain direction-of-arrival information, whilst being highly computationally and parametrically efficient than standard 1D/2D convolution. A dense sound event proposal map is then constructed to handle the challenges of predicting events with large varying temporal duration. Accompanying the dense proposal map are a temporal overlapness map and a motion smoothness map that measure a proposal's confidence to be an event from temporal detection accuracy and movement consistency perspective. Involving the two maps guarantees SoundDet to be trained in a spatio-temporally unified manner. Experimental results on the public DCASE dataset show the advantage of SoundDet on both segment-based and our newly proposed event-based evaluation system.
\end{abstract}

\section{Introduction}

Acoustic source detection, classification and localization is a key task for a wide variety of applications, such as smart speakers/assistants, interactive robotics and scene visualization. Broadly, a microphone array (typically with four microphones) receives sound events from the environment, which are corrupted by background noise. The task is to firstly detect events, classify them, and finally estimate the physical location or direction-of-arrival angle (DoA). 

Early work has demonstrated how deep learning techniques can be used to provide good performance for sound event detection and localization. However, the majority of these works treat event detection and localization as separate problems\cite{seld_dcase19}\cite{polyphonic_seld_icassp20}. For example, a framewise classifier and regressor are used separately \cite{doa_esti_multi_source}\cite{sound_loc_study} to obtain event class and spatial location. In addition, they also rely heavily on handcrafted pre-processing to convert the raw waveforms into time-frequency representations\,(\textit{e.g.} log-mel spectrograms, gcc-phat\cite{gcc_phat}) that are more amenable to process with mature 2-D CNN networks like ResNet\cite{resnet18} followed by LSTM\cite{LSTM} or GRU\cite{GRU} network for handling temporal dependencies. Lastly, sound event detection is typically achieved following a segment-based approach (\textit{e.g.} cut the waveform into second-long snippet) which inevitably leads to incomplete or partial detection that span segment boundaries.

The relatively unexplored area becomes more challenging when it considers polyphonic event detection where events with different DoAs overlap temporally. In addition, rather than necessarily being stationary, these events undergo a motion. This increased realism leads to poor performance or even incapability of existing methods\cite{joint-measure}\cite{seld_dcase19} that make a strong assumption that only a single and stationary event exists at a time slot.

In this paper, we rethink the polyphonic sound event detection and localization problem. We draw inspiration from the success of object detection in 2-D images and 3-D point clouds, and think of an event as being analogous to a ``sound-object" that has a specific location (spatial and temporal onset/offset), size (duration) and class (semantic category). 

Unlike computer vision where the primary challenge is occlusion of overlapping objects leading to partial views, the physics of sound are such that sound-objects superimpose and additively mix with one another. Sound objects thus have to rely on frequency-selective and phase-sensitive approaches so as to be distinguished from background or ambient noise. Rather than detecting in several separate and independent steps, a more desirable event detector needs to be end-to-end that directly consumes raw waveform and outputs predictions, whilst generates minimal computation cost. To this end, we propose a unified and light weight polyphonic sound event detection and localization framework, dubbed SoundDet, that naturally meets these requirements. 

To the best of our knowledge, SoundDet is the first approach to directly consume the raw waveform and not rely on any spectral transformation pre-processing. To achieve this, we propose to learn a set of channel correlation aware and frequency sensitive bank of filters, these parametric filters naturally captures the phase difference between channels and frequency bandpass for different sound events. Comparing with standard 1D/2D convolution, learning such filter bank requires minimal computation cost and parameters, making it ideal for processing the raw waveform with large sampling rates. SoundDet takes a novel view and adopts a dense event proposal strategy to directly estimate an event's location, size and category information, unlike current approaches that simply treat it as a segment-wise prediction problem. To better model a sound event's completeness and continuity, we introduce concept like motion smoothness, boundary sensitive temporal overlapness.

To comprehensively evaluate the performance, we further propose an event-based evaluation system. Unlike previous segment-based evaluation, event-based metric takes sound event's confidence score, length and motion consistency into account and accumulates performance under different score threshold. We embrace 2D image object detection metric\cite{coco_dataset} and propose to calculate average precision/recall, mean average precision (mAP), mean average recall (mAR) for each category. Experiment on 14 sound category DoA estimation DCASE dataset\cite{dcase2020} shows the advantage and efficacy of SoundDet. 

In summary, our main contributions are three fold: \textbf{First}, We propose the first framework for polyphonic moving sound event detection and localization that directly consumes raw waveform and estimate sound event from ``event-object" perspective. \textbf{Second}, We propose a novel phase and frequency sensitive MaxCorr filter bank that is parameter and computation lightweight to process raw waveform. \textbf{Third}, An event-based evaluation system that avoids arbitrarily setting threshold is proposed to better evaluate the framework.

\section{Related Work}

Existing approaches\cite{seld_dcase19}\cite{eventness_detect}\cite{polyphonic_seld_icassp20} on sound event detection and localization heavily rely on 2D convolutional neural networks to process the large, multichannel waveforms. In order to use 2D CNN frameworks such as ResNet\cite{resnet18_seld}, the 1D waveforms are converted 2D time-frequency representations by using hard-coded methods. Typical method include short-time Fourier Transforms\,(FFT)\cite{kothinti2019joint,krause2019arborescent},  Generalized Cross-Correlation (GCC-PHAT)\cite{adavanne2016sound,gcc_phat},
Mel-Spectrograms\cite{MFCC,cakir2016filterbank,hayashi2017duration}, Log-Mel Spectrograms\cite{adavanne2016sound,xia2019multi} and pitch estimators\cite{adavanne2016sound}. Contrary to these methods, SoundDet is an end-to-end and unified framework, it directly consumes raw waveform and jointly predicts a potential event by fully considering it as a holistic ``sound-object", embedded in time and space. It does not require any hand-selection and tweaking of preprocessing algorithms. Moreover, existing approaches usually chain 2D CNNs and recurrent networks (\textit{i.e.} GRU\cite{GRU}, LSTM\cite{LSTM}) to incorporate temporal information. SoundDet consists of light-weight CNNs and hence has few parameters and is computationally efficient.

Fair and comprehensive evaluation of the sound event detection and localization is still an under-defined problem. Existing methods adopt segment-based evaluation, in which they segment a sound track into small chunks (\textit{i.e.} 1s) and compare the result just within a segment, ignoring a sound event's duration and continuity. In addition, the segment-based evaluation treats each detected sound event as either true positive or false positive by arbitrarily setting a threshold. It does not represent the performance under various thresholds. We draw inspiration from object detection in 2D images\cite{coco_dataset}\cite{faster_rcnn}\cite{SSD} and  propose a new event-based evaluation metric, which instead computes mean average precision\,(mAP) and mean average recall (mAR) for each category accumulatively.

\section{SoundDet}

\subsection{Problem Formulation}
We have a multi-channel sound recording\,(aka sound waveform) $W_c$ recorded by one microphone array station or multiple microphone array stations at a fixed sampling rate, $W_c$ contains a sound event set\,\footnote{In this paper, a sound event is restricted to be a temporally and spatially continuous and semantically meaningful. Two events with exactly the same temporal location belong to different class.} $E=\{e_i=(t_{s,i}, t_{e,i}, l_i, c_i)\}_{i=1}^{N}$, where $t_{s,i}$, $t_{e,i}$ are the $i$-th sound event's start time and end time, respectively, $c_i$ indicates the sound event's semantic information\,(\textit{i.e.} category, baby cry, phone ring). $l_i$ indicates the sound event's spatial location, it can either be a fixed location if the sound event is stationary or a spatial trajectory if it undergoes a motion. The task is to learn a model $\mathcal{F}$ to accurately recover these sound events directly from raw sound waveform,

\begin{equation}
    E = \mathcal{F}(W_c)
\end{equation}

These sound events are polyphonic, which means they may overlap in the time dimension.

\subsection{MaxCorr Band-Pass Filter Bank and Backbone}

\begin{figure*}
    \centering
    \includegraphics[width=0.97\textwidth]{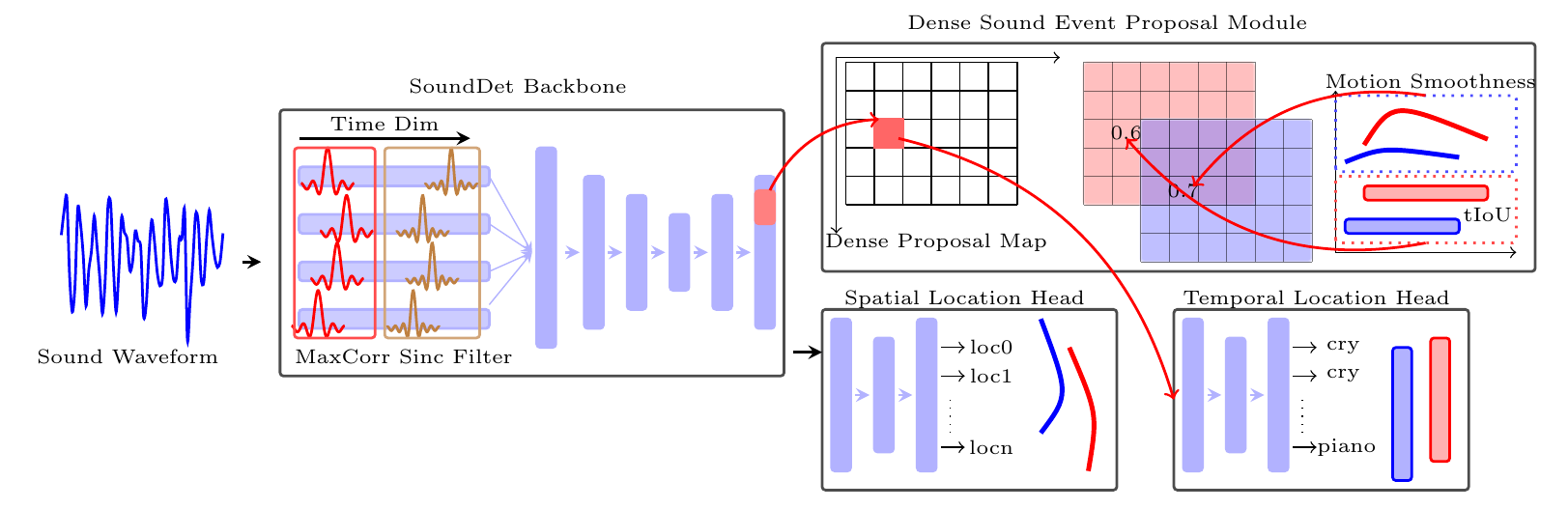}
    \caption{SoundDet pipeline: the input multi-channel waveform is fed to a backbone neural network to learn framewise representations. The backbone consists of MaxCorr filter bank and a encoder-decoder like network. A spatial location head and a dense event proposal generation head connect the backbone network in parallel. The spatial location head predicts per-class spatial location in a framewise manner. The dense proposal head gives all possible event proposals, each one associates with an event-wise feature, certain start/end time. The dense proposal head maintains two score maps, one map measures proposals' temporal overlap with group truth and the other represents proposals' motion smoothness, which directly comes from spatial location head. The two maps unifies spatial location head and temporal location head so that the whole framework can be jointly trained in a unified manner.}
    \label{fig:pipeline}
\end{figure*}

Sound waveform usually has high temporal resolution, a typical 1 minute recording of 30kHz has 1.8 million data points. This leads to unbearable computation burden\,(FLOPs) for standard 1D/2D convolution.
To counteract the large data point size issue and further to enforce the neural network to learn meaningful sound event spatio-temporal representation, we propose to learn a set of rectangular band-pass filters in frequency domain, each of which contains a lower learnable frequency cuttoff $f_1$ and a higher learnable frequency cutoff $f_2$. After converting the rectangular band-pass filter to time domain, we obtain $sinc$ convolution kernels $k$,

\begin{equation}
k[n, f_1, f_2] = 2f_2 sinc(2\pi f_2 n) - 2f_1 sinc(2\pi f_1 n)
\end{equation}

where $sinc(x) = sin(x)/x$. Note that the band-pass filter is a parametric filter that the learnable parameter number is independent of the kernel length. Moreover, the kernel length is usually much larger than standard 1D/2D convolution kernel length. A typical band-pass filter kernel length can be set as a large number\,(in our case 481), whereas the standard 1D/2D convolution kernel size is usually 3 or 5. These two properties guarantee these band-pass filters are highly efficient at both FLOPs computation and model parameter number. The SincNet filter has been successfully applied for speech recognition\cite{SincNet}. 

\begin{figure}[t]
    \centering
    \includegraphics[width=0.9\linewidth]{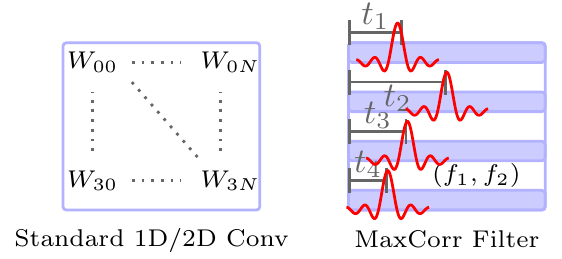}
    \caption{Visual comparison of standard 1D/2D convolution with our proposed MaxCorr filter. While the learnable parameters of standard 1D/2D convolution equals to filter size\,(left side image), MaxCorr filter is parametric and introduces much fewer learnable parameters (right side image). For four channel waveform inputs, each MaxCorr filter merely has six learnable parameters (two frequencies and four time shifts).}
    \label{fig:max_corr}
\end{figure}

To further enforce the above designed filter banks to be able to dynamically decouple the sound event's temporal information and spatial information. We further expand the sinc filter to multiple channels to model the internal correlation between different channels. We draw inspiration from canonical time difference of arrival (TDoA) estimation to explicitly encode sound events' geometrical information recorded by microphone array. TDoA is a well-established technique for sound source geolocalization. It exploits the arrival time difference among microphones to recover the sound event spatial information\,(\textit{i.e.} DoA or physical location) by computing the maximum time delay in either frequency domain or time domain\cite{time_domain_gcc}. We involve TDoA estimation into our band-pass filter bank design and propose a new filter bank $g(\cdot)$ that is capable of learning the correlation between different channels. Learning such filter bank helps the neural network to dynamically learn how to integrate between-channel correlation and event frequency band pass to best recover sound events, we thus call it MaxCorr filter bank,

\begin{equation}
    g(n, f_1,f_2, t_1,\cdots, t_c) = [k[n+t_1], \cdots, k[n+t_c]]
    \label{maxcorr-filter}
\end{equation}

where $t_i$ is the $i$-th channel time shift parameter, $f_1$, $f_2$, $t_1, \cdots, t_c$ are learnable parameters and in our case $c=4$ as there are four channels. To make the time shift parameter to be differentiable, we adopt sigmoid like soft truncation to approximate the round operation, like \cite{look_closer} do. Please see Fig.\ref{fig:max_corr} for MaxCorr filter illustration and its comparison with standard convolution.

Following the MaxCorr filters, we add an encoder-decoder like 1D convolution network to learn framewise representations. In the encoder, we gradually reduce the temporal length but the increase the filter number, resulting in a compressed representation. In the decoder, it goes the opposite way and finally forms a ``bottleneck"-like backbone neural network. To mitigate the model training degradation dilemma, we add skip connection between the encoder and decoder. The MaxCorr filter bank and encoder-decoder like neural network together constitute the Backbone neural network, which learns a framewise representation $[f_1, f_2, \cdots, f_N]$, where each representation's temporal duration equals to ground truth labelling resolution, like 100ms.

\subsection{Dense Sound Event Proposal}

Given the representation learned by the backbone, the next step is to generate sound event proposal.
Potential sound events freely span in the temporal and spatial space, which means sound events are largely varying in their temporal length and spatial location. An efficient sound event proposal generation module thus should be: 1. able to handle sound events with large varying temporal length and spatial location. 2. computationally efficient enough\,(\textit{i.e.} generating potential event proposals at parallel). To this end, we propose a dense sound event proposal generation module, which organizes all potential sound events into a compact organization and all the potential events can be generated in parallelization.

Specifically, we compactly organize all potential sound events into a matrix-like representation $\mathcal{M}$, where the row indicates the sound event's temporal length and the column represents its start time. Each cell $C_{i,j}$ in $\mathcal{M}$ corresponds to a sound event with certain start time $j$ and end time $i+j$. A careful selection of the size of matrix $\mathcal{M}$ guarantees all potential sound events have their unique ``cell" in $\mathcal{M}$. The matrix-like compact organization enables fast computation due to matrix parallelization computation supported by most GPU/CPU engines. Similar idea has been used for video-based activity recognition\cite{bsn}.

Accompanying the dense proposal map are two score maps of the same size: $\mathcal{M}_t$ measuring a sound event proposal's temporal overlap with ground truth, and $\mathcal{M}_s$ sound event proposal motion smoothness map. The two score maps jointly help to measure sound event proposals' confidence to be a true positive sound event. Specifically, the potential proposal indicated by cell $C_{i,j}$ is associated with a event wise feature representation $f_{ij}$, a temporal overlap confidence score $s_{i,j}^t$ from $\mathcal{M}_t$ and a motion smoothness score $s_{i,j}^s$ from $\mathcal{M}_s$. $f_{i,j}$ derives from the averaging features learned by Backbone network spanning its temporal duration $[t_j, t_{i+j}]$, $f_{i,j} = \frac{1}{i} \sum_{k = j}^{i+j} f_k$.

\begin{figure}[t]
    \centering
    \includegraphics[width=0.97\linewidth]{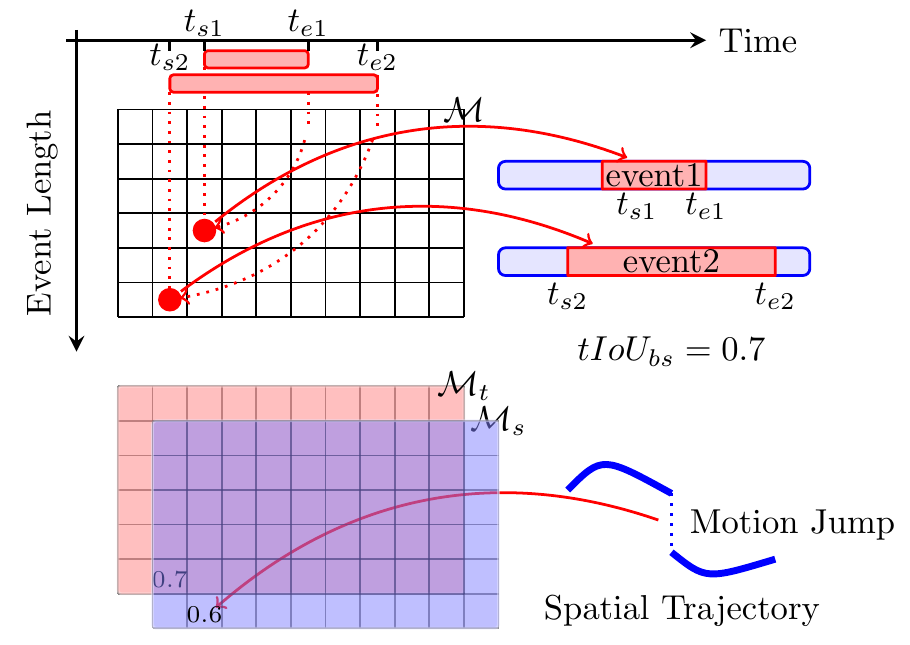}
    \caption{Dense sound event proposal module. A matrix $\mathcal{M}$ is created to densely represent all potential sound events, each cell of the matrix corresponds to a sound event with certain start time and end time. The matrix maintains two score maps $\mathcal{M}_t$ and $\mathcal{M}_s$ of the same size. While $\mathcal{M}_t$ records proposals' temporal overlap with ground truth, $\mathcal{M}_s$ measures proposals' motion smoothness. The two score maps jointly represent a proposal's likeliness to be a true positive event. The motion jump, for example, largely violates the motion smoothness rule, so it is highly unlikely that the proposal is a true event.}
    \label{fig:dense_proposal}
\end{figure}

\subsection{Temporal Overlapness Score Map}

Temporal overlap score map $\mathcal{M}_t$ measures two sound events' temporal overlap degree. Here we adopt temporal intersection over union (tIoU). Given two sound events $s_1 = [t_{s1}, t_{e1}]$ at cell $\mathcal{M}_{i,j}$ and its ground truth temporal location $s_2 = [t_{s2}, t_{e2}]$, their tIoU at $\mathcal{M}_t^{i,j}$ is defined as,

\begin{equation}
    tIoU = \frac{ max\{0, min(t_{e1}, t_{e2}) - max(t_{s1}, t_{s2})\}}{max(t_{e1}, t_{e2}) - min(t_{s1}, t_{s2})}
\end{equation}

The higher of tIoU, the more likely the proposal at $C_{i,j}$ is a true positive proposal. Note that $\mathcal{M}_t$ is class-agnostic and its responsibility is to generate all potential event proposals\,(also can be thought as eventness proposal). In training, we train a $\mathcal{M}_t$ map regressor to learn the mapping from event wise feature representation to tIoU score. During test, we partially use the regressor to generate raw sound event proposals by choosing the large scores. To further enhance to regressor to fast localize a sound event, we propose a boundary sensitive tIoU $tIoU_{bs}$ which explicitly penalizes even a small temporal location alteration by multiplying an exponential term,

\begin{equation}
    tIoU_{bs} = tIoU \cdot e^{-w\cdot (1-tIoU)}
\end{equation}

where $w$ is a decay weight controlling tIoU decay rate, which we set 2. Our observation is that as a sound event temporal length increases, slight temporal alteration results in small tIoU score change. For example, given the ground truth sound event resides in frames [20,80] and the proposal is [22,82], the original tIoU will be very high (0.94 in this case) as  there  is  a  high  degree  of  overlap. This  does  not  give sufficient steering to the network to correctly align the event boundaries. Under our new metric, however, $tIoU_{bs} = 0.83$, the misalignment is prominently magnified so that the network receives enough clue to improve its localization capability.

\subsection{Spatial Motion Smoothness Map}
An independent sound event's spatial trajectory should be consistent and smooth. In other words, there should be no abrupt ``jump" along the sound event's trajectory, regardless of its motion status. We call this property as motion smoothness. Specifically, we model motion smoothness as the maximum neighboring location displacement along the sound event's motion path. Mathematically, for the sound event $C_{i,j}$'s sequential spatial location $\{l_0, \cdots, l_{i}\}$, the motion smoothness is defined as the maximum neighboring spatial location displacement $d$,

\begin{equation}
    \mathcal{M}_s^{(i,j)} = \max\{d(l_k - l_{k+1})\}_{k=0}^{i-1}
    \label{eq_motion_smoothness}
\end{equation}

$d(\cdot)$ is modelled as squared Euclidean distance. Small motion smoothness score serves as a strong indicator of the existence of a potential complete sound event because an inaccurate sound event localization inevitably introduces large ``motion jump". We hereby jointly use $\mathcal{M}_s$ and $\mathcal{M}_t$ to efficiently generate proposals. The introduction of motion smoothness map has two main advantages. First, it integrates temporal/spatial head so that the whole framework can be trained in a unified manner. Second, it explicitly adds temporal regularization to spatial location head, because spatial location head alone is a framewise regressor and no temporal priors are involved. Please see Fig.\ref{fig:dense_proposal} for dense event proposal illustration and its connection with $\mathcal{M}_t$ and $\mathcal{M}_s$. 

\subsection{Framewise Spatial Location Regression}

The final component of SoundDet is the spatial location head. As the the sound source physical location or DoA can change over the duration of an event, the spatial location is computed in a framewise manner, using the framewise features extracted by the backbone network. We regress per-class spatial location for each frame and all inactive sound classes’ spatial location is set 0. 

\subsection{Training Pipeline}

The overall SoundDet pipeline is shown in Fig.\ref{fig:pipeline}. The multi-channel raw waveform input is fed to the backbone neural network to efficiently learn framewise representation. The spatial location recovery head and dense event proposal generation head directly build on the learned framewise representation to learn framewise per-class sound event spatial locations and densely generate event proposals\,(per-event representation), respectively. The densely generated event proposals are fed to temporal recovery head to learn proposals' category label.

In practice, the temporal location head builds on event wise feature. It consists of three sub-heads: a multi-label classification head deciding the proposal's category label\,($head1$); A binary classification head deciding whether a proposal is a foreground or background event\,(we can call it eventness prediction for better understanding). A tIoU regression head to learn to regress tIoU score based on the event wise feature input. The three heads consist of several full connection layers (FC). For multi-label classification and binary classification, we adopt the standard cross entropy loss. For the tIoU regression head, we adopt mean squared error\,(MSE) loss. The computation of $\mathcal{M}_s$ requires no specific learning process because it directly derives from spatial location head, MSE loss is adopted again to reinforce event's smoothness.

\textbf{Data Imbalance} The dense sound event proposal map presented above covers all possible events, it inevitably leads to the data imbalance problem because most pre-generated proposals are negative and just very few proposals are positive. We adopt two strategies to mitigate this problem, 1) increase positive proposal number, in which we set a tIoU threshold $t_d$ and any event with the tIoU larger than the threshold is treated as positive proposal. 2) negative proposal random dropout, with the remaining negative samples, we randomly drop part of them with a probability $p_d$. By carefully setting $t_d$ and $p_d$ value, we can roughly keep positive and negative ratio to be 1:1.

\subsection{Inference}
\label{sec_inference}
During inference, the calculated $\mathcal{M}_s$ and $\mathcal{M}_t$ are jointly used to generate sound event proposals. The sound event proposal at cell $C_{i,j}$ is treated as positive proposal only if it passes two test: 1) $\mathcal{M}_s^{i,j} > d_s$ so that it satisfies motion smoothness rule; 2) $\mathcal{M}_t^{i,j} > d_t$ so that it resides in the designated temporal location $(t_j, t_{i+1})$ with a high confidence. $d_s$ and $d_t$ are predefined spatial motion smoothness and temporal overlapness threshold. Once it passes the two tests, the final event detection score derives from the multiplication of its motion smoothness score, temporal overlapness score and the corresponding multilabel classification score\,(($head1$) discussed earlier). The class label comes from the same multilabel classification head and the spatial location derives from spatial location head.

\begin{figure}[t]
    \centering
    \includegraphics[width=0.9\linewidth]{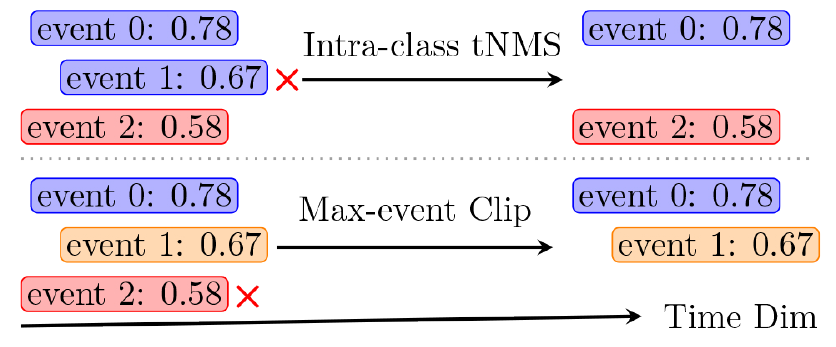}
    \caption{SoundDet postprocess illustration: we apply inra-class NMS to suppress sound events of the same classes along the temporal axis. Given multiple different temporally-overlapping sound events, max-event clip process is applied to clip untruthful events. Different colors indicate different sound event classes.}
    \label{fig:post_process}
\end{figure}

\textbf{Event Refinement} The raw event candidates obtained above overlap in the temporal dimension. A post-processing is thus required to remove the redundancy (see Fig.\ref{fig:post_process}). The post-processing consists of two main parts: intra-class temporal non-maximum suppression (tNMS) and max-event clip. In intra-class tNMS, we compare within the same event class, for two events of the same class with temporal overlap (tNMS) above a predefined threshold, we delete the event with smaller confidence score. In max-event clip, we restrain the maximum number events that can happen at the same time. Max-event clip is a class-agnostic operation, for a set of events that happen at the same time, we sort them according the confidence score $s_i$ in descending order and only keep the top-M events. $M$ is a predefined max-event threshold. The post-processing is an iterative process, after which we get the final $K$ detected sound events from $N$ rw proposals $K\leqslant N$.

\section{Experiment}

\begin{table*}[t]
    \centering
    \small
    \caption{Segment-based evaluation result. We report result for different event length: All (overall), Small (0-2s), Medium (2-7s) and large ($> 7$s.) SoundDet is more advantageous at predicting longer sound events than EIN\cite{ein_v2}}
    \begin{tabular}{|p{3.3cm}<{\centering}|p{0.5cm}<{\centering}p{0.4cm}<{\centering}p{0.4cm}<{\centering}p{0.4cm}<{\centering}|p{0.4cm}<{\centering}p{0.4cm}<{\centering}p{0.4cm}<{\centering}p{0.4cm}<{\centering}|p{0.4cm}<{\centering}p{0.4cm}<{\centering}p{0.4cm}<{\centering}p{0.4cm}<{\centering}|p{0.45cm}<{\centering}p{0.4cm}<{\centering}p{0.4cm}<{\centering}p{0.4cm}<{\centering}|}
        \hline
        \multirow{2}{*}{Methods} &\multicolumn{4}{c|}{$ER_{20^\circ}(\downarrow)$} & \multicolumn{4}{c|}{$F_{20^\circ}(\uparrow)$} & \multicolumn{4}{c|}{$LE_{CD}(\downarrow)$($^\circ$)} & \multicolumn{4}{c|}{$LR_{CD}(\uparrow)$} \\
        \cline{2-17}
        &All&Sma&Mid&Lar &  All&Sma&Med&Lar & All&Sma&Med&Lar & All&Sma&Med&Lar\\
        \hline
        SELDnet(foa)\shortcites{seld_dcase19} & 0.63 & 0.64 & 0.63 & 0.79 & 0.46 & 0.48 & 0.48 & 0.42 & 23.1 & 22.8 & 23.2 & 20.5 & 0.69 & 0.70 & 0.70 & 0.64 \\
        SELDnet(mic)\shortcites{seld_dcase19} & 0.66 & 0.68 & 0.66 & 0.79 & 0.43 & 0.44 & 0.45 & 0.42 & 24.2 & 23.9 & 23.6 & 22.8 & 0.66 & 0.65 & 0.67 & 0.66 \\
        EIN\cite{ein_v2} & \textbf{0.25} & \textbf{0.30} & 0.25 & 0.29 & \textbf{0.82} & \textbf{0.80} & 0.82 & 0.83 & \textbf{8.0} & \textbf{8.1} & 8.0 & \textbf{8.5} & \textbf{0.86} & \textbf{0.85} & 0.86 & 0.84 \\
        \hline
        SoundDet\_backbone(foa) & 0.74 & 0.79 & 0.75 & 0.74 & 0.38 & 0.37 & 0.40 & 0.38 & 21.7 & 22.2 & 16.7 & 21.7 & 0.57 & 0.57 & 0.51 & 0.57 \\
        SoundDet\_backbone(mic) & 0.80 & 0.85 & 0.82 & 0.90 & 0.30 & 0.29 & 0.31 & 0.34 & 28.6 & 29.5 & 28.6 & 21.0 & 0.56 & 0.56 & 0.57 & 0.54 \\
        SoundDet\_nomaxcorr(foa) & 0.90 & 0.91 & 0.90 & 0.90 & 0.05 & 0.05 & 0.06 & 0.06 & 79.4 & 79.3 & 79.8 & 79.0 & 0.50 & 0.51 & 0.53 & 0.48 \\
        SoundDet\_nomaxcorr(mic) & 0.90 & 0.95 & 0.95 & 0.90 & 0.04 & 0.04 & 0.04 & 0.04 & 87.3 & 85.9 & 87.4 & 91.3 & 0.48 & 0.48 & 0.49 & 0.49 \\
        \hline
        SoundDet\_nomots(foa) & 0.31 & 0.37 & 0.38 & 0.35 & 0.77 & 0.72 & 0.74 & 0.72 & 9.0 & 10.1 & 9.8 & 7.4 & 0.77 & 0.72 & 0.74 & 0.76 \\
        SoundDet(foa) & \textbf{0.25} & 0.31 & \textbf{0.24} & \textbf{0.26} & 0.81 & 0.79 & \textbf{0.83} & \textbf{0.86} & 8.3 & 8.5 & \textbf{7.7} & 8.1 & 0.82 & 0.80 & \textbf{0.89} & \textbf{0.85} \\
        \hline
    \end{tabular}
    \label{table_segment_based_eval_new}
\end{table*}

SoundDet is capable of estimating 2D/3D DoA, physical location, distance and motion of various sound events in polyphonic and moving scenario. In this experiment, we focus on indoor 3D DoA estimation tasks. 

\textbf{Dataset} We evaluate SoundDet on TAU-NIGENS DCASE sound event detection and localization (SELD)\cite{dcase2020} dataset. It is an indoor synthetic sound recording collected by an Eigenmike spherical microphone array and available in two data formats: first-order ambisonics (FOA) and tetrahedral microphone array (MIC). MIC indicates four microphones with different orientations, usually in spherical coordinates.  FoA is known as B-format, consisting of omni-directional and  $x, y, z$ direction components. The possible azimuths of the sound recording span the whole range $[-180^\circ, 180^\circ)$ and the elevations lie in range $[-45^\circ, 45^\circ]$. Recorded sound events are either stationary or moving, and a maximum of two sound events overlap spatially and temporally. In total, 14 sound event categories are generated: alarm, crying baby, crash, barking dog, running engine, female scream, female speech, burning fire, footsteps, knocking on door, male scream, male speech, ringing phone and piano. Each sound recording is 1 minute long with sampling rate 24~kHz. For more details about this dataset, please refer to \cite{dcase2020}. There are 8 folds recordings, each of which has 100 1-min $.wav$ format recording. We follow the official splits and use 1-6 folds for train and the remaining 7-8 folds (200 1-min recordings) for test.

\subsection{Evaluation Metrics}

\textbf{Segment-based Metric} is the standard evaluation metric for existing methods \cite{seld_dcase19},\cite{doa_esti_multi_source},\cite{joint-measure}. It compares prediction and ground truth within non-overlapping short temporal segments\,(one sec.) and evaluates each frame's prediction result by jointly considering its predicted class and spatial location. For event detection evaluation, it calculates $F1$-score ($F_{\leq T^\circ}$) and Error Rate\,($ER_{\leq T^\circ}$), where a true positive prediction has to be within a spatial distance threshold with the corresponding ground truth spatial location\,(typically with an angular threshold $T = 20^\circ$). For event localization evaluation, we compute class dependent localization error $LE_{CD}$ which measures the average location distance between prediction and ground truth, and localization recall $LR_{CD}$ measures the ratio of how many such localizations were estimated within a class. For details, please refer to \cite{joint-measure}.   

\textbf{Event-based Metric} draws inspiration from evaluation on 2D image object detection. It treats a sound event as an independent instance with specific start time, end time, framewise spatial location and confidence score belonging to one class\,(see Sec.\,\ref{sec_inference}). Rather than simply binarizing a detection as false or true detection with an arbitrary threshold, event-based metric comprehensively evaluates performance under various event confidence score and tIoU thresholds, calculating average precision (AP) and average recall (AR) for each class separately. Finally, mean average precision (mAP) and mean average recall\,(mAR) is accumulated by averaging AP and AR throughout classes respectively. 

Specifically, given $K$ detected sound events $E_p = \{t_{s,i}, t_{e,i}, l_i, c_i, s_i\}_{i=1}^{K}$ and $N$ ground truth sound events $E_{gt} = \{t_{s,i}, t_{e,i}, l_i, c_i\}_{i=1}^{N}$ of the same class, we compute the average precision and average recall under tIoU in range $[0.1, 0.05, 1.0]$, where $0.05$ is the stepsize. For a particular tIoU, the AP and AP is computed by averaging precision scores and recall scores obtained under various confidence score in range $[0.1, 0.05, 1.0]$. Please note that we don't take any predefined tIoU and confidence score threshold because arbitrary chosen thresholds inevitably introduces human bias e.g. the choice of an entirely arbitrary angular threshold $T = 20^\circ$ in the segment based approach. Our proposed event-based metric instead provides a more objective and comprehensive metric.

\subsection{Comparing Methods} 

We compare SoundDet with two existing methods: SELDNet\cite{seld_dcase19} and EIN\cite{ein_v2}. Since we focus on polyphonic and moving scenario, other relevant methods\cite{polyphonic_seld_icassp20}\cite{two-stage} that merely work on stationary and monophonic sound events are not discussed here. SELDNet\cite{seld_dcase19} jointly trains sound event detection and localization with CRNN network. It extracts Log-mel, GCC-PHAT, Intensity features as neural network input. Bidirectional GRU network is involved to model temporal dependency. SELDNet is treated as the baseline. EIN\cite{ein_v2} is a very recent work which models sound event detection and localization with two identical but separate neural networks as a multi-task learning format. It uses a log-mel spectrogram as input for event detection, and a GCC-PHAT approach for DoA estimation. The two parallel networks are independent but with soft parameter sharing. In addition, multi-head self-attention\cite{attention} is applied. EIN has large parameter size and is currently the state of the art approach under segment-based evaluation metric.

The above two methods generate framewise predictions, in which each frame is associated with class classification score and predicted DoA value. In order to generate a score that metrics event classification score and DoA closeness between predicted location and ground truth location, we propose to integrate mean classification score $s_c$ and mean DoA Euclidean distance $d_{DoA}$ together in the formula $s=s_c\cdot e^{-d_{DoA}}$. For fair comparison, we don't involve any data augmentation methods. We report the result for both FOA and MIC data format if possible.

Within SoundDet, we want to answer four main questions: 1. If our proposed MaxCorr band-pass filters is capable of learning useful representation for sound event spatio-temporal information recovery? 2. Event length impact on SoundDet and other methods. 3. Efficiency comparison in terms of parameters and inference time. 4. Quantitative comparison between SoundDet and other methods. To this end, we test four SoundDet variants: backbone only SoundDet with MaxCorr filter\,(SoundDet\_backbone) and without MaxCorr filter\,(SoundDet\_nomaxcorr). SoundDet without motion smoothness\,(SoundDet\_nomots). Moreover, in addition to report the overall metric, we further divide the events into Small\,(0-2s), Medium\,(2s-7s) and Large\,($>$7s), three categories and report result for them separately.

SoundDet network architecture is shown Table\,\ref{SoundDet_network}. Please note that the network might need to be slightly changed to process waveform with different input length, sample rate and label interval length. $H$ and $W$ indicate dense proposal map height and width, respectively, in our experiment $H=60$ and $W=60$, they can be directly modified to fit other input waveform temporal length or ground truth labelling resolution.

\textbf{Training Detail} We adopt multi-stage training strategy. First, we train the SoundDet backbone network in a framewise DoA regression and multilabel classification manner, like SELDNet\cite{seld_dcase19} and EIN\cite{ein_v2} do. Training SoundDet backbone neural network at first provides two advantages: 1) it helps to test if our proposed MaxCorr filter bank helps to learn essential representation for sound event temporal detection and spatial localization\,(as we compare in the experiment), 2) it guarantees to learn a reasonable framewise representation first so that the later joint training\,(combine the backbone and two heads) converges much faster. For the backbone training, we use SGD optimizer with an initial learning rate 0.5, the learning rate decays every 30 epochs with decay rate 0.7. With the pretrained Backbone network, we continue to train the whole SoundDet network with the same optimizer.

\begin{figure}[t]
    \centering
    \small
    \includegraphics[width=0.85\linewidth]{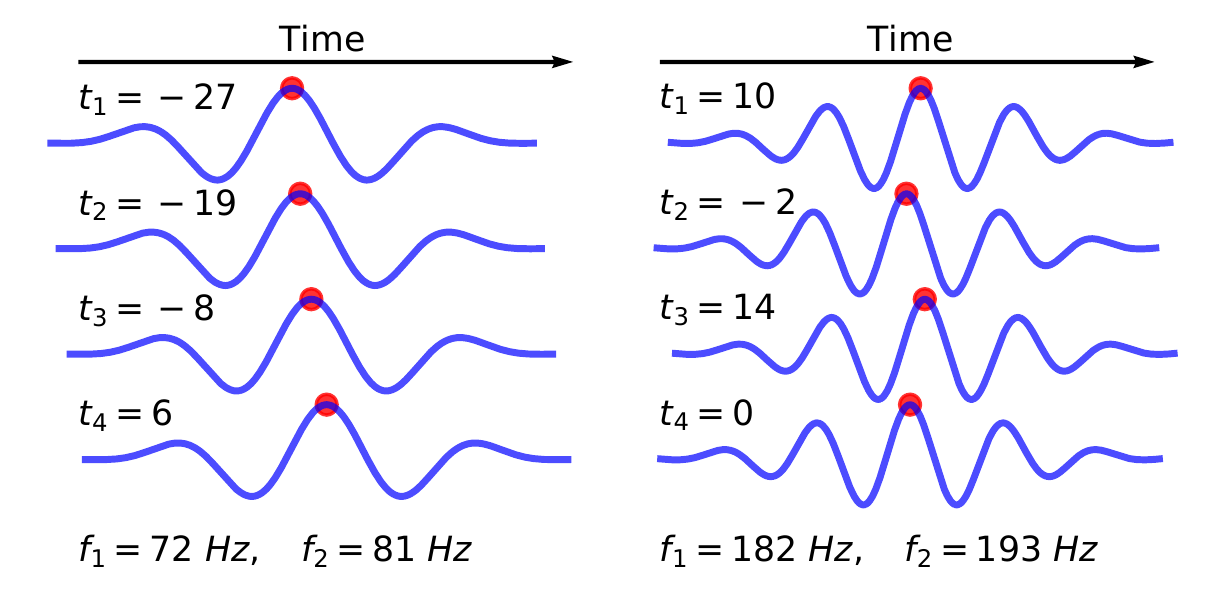}
    \caption{Two learned MaxCorr filters, they are controlled by different frequency cutoffs and four time shifts. The red dots are the filter symmetric point.}
    \label{fig:learned_maxcorr_filter}
\end{figure}

\begin{figure}[t]
    \centering
    \includegraphics[width=0.85\linewidth]{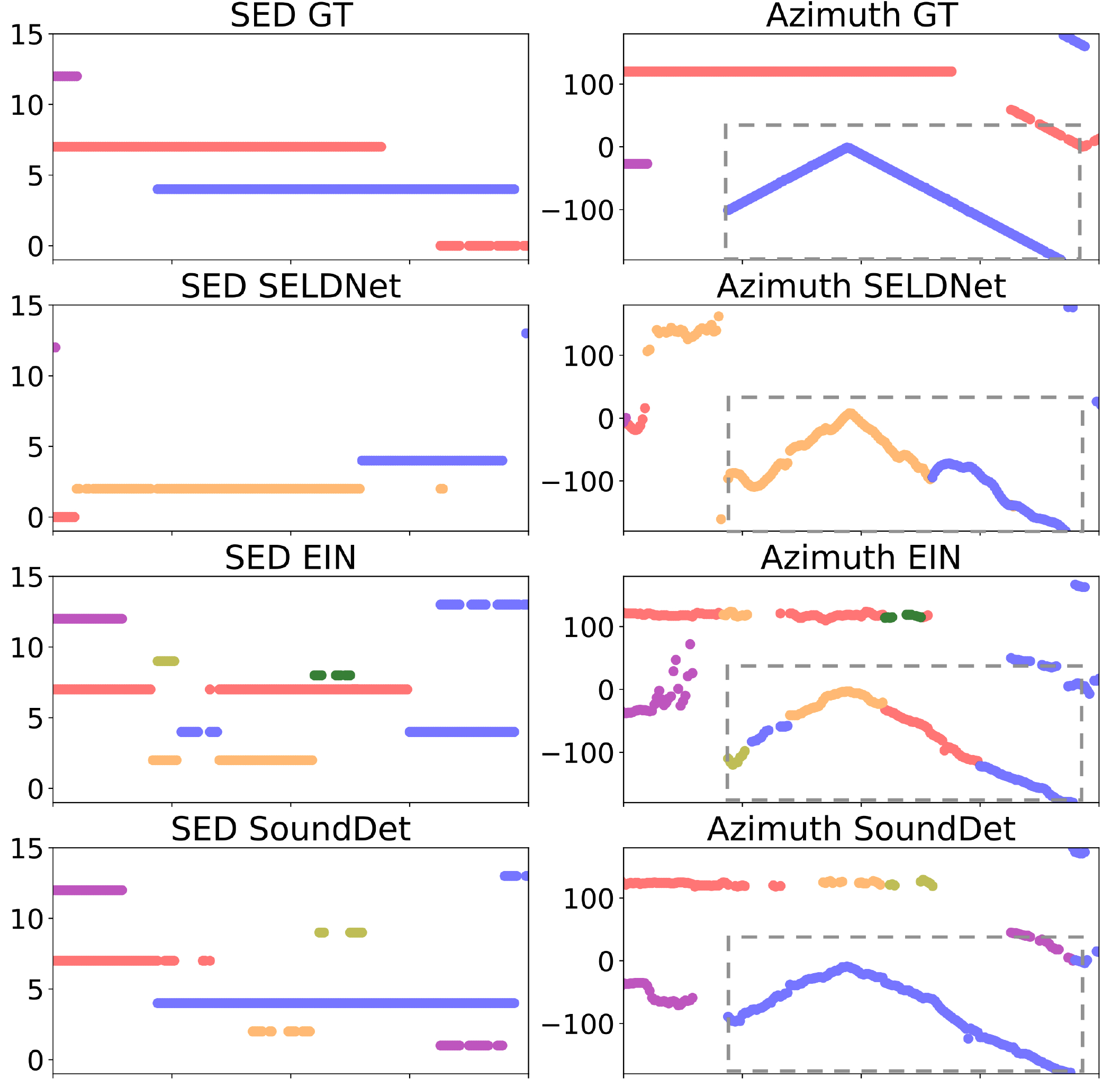}
    \caption{Qualitative comparison. We visualize detected sound events' temporal location\,(SED, left side) and Azimuth-only DoA (right side). While framewise based methods\,(SELDNet and EIN) produce discrete and segmented event, SoundDet maximally keeps an event's completeness and continuity. Pay attention to dotted grey box and different color indicates different sound event class.}
    \label{qualitative_result}
\end{figure}

\subsection{Experiment Results}

\begin{table*}[h]
    \small
    \centering
    \caption{Event-based evaluation result, model parameter number and input feature. We report mAP/mAR under different event temporal length threshold. The “Input” column labels are: 0. Raw waveform, 1. Log-Mel, 2. GCC-PHAT, 3. Intensity.}
    \begin{tabular}{|c|c|c|cccc|cccc|}
    \hline
    \multirow{2}{*}{Methods} & \multirow{2}{*}{Params}&\multirow{2}{*}{Input} &\multicolumn{4}{c|}{mAP$(\uparrow)$} & \multicolumn{4}{c|}{mAR$(\uparrow)$} \\
    \cline{4-11}
    &&&Overall&Small&Medium&Large&Overall&Small&Medium&Large\\
    \hline
    SELDNet(foa)& 0.5M&1,3&0.087 & 0.038 & 0.092 & 0.157 & 0.152 & 0.079 & 0.096 & 0.097\\
    SELDNet(mic)& 0.5M&1,2&0.079 & 0.035 & 0.081 & 0.143 & 0.140 & 0.067 & 0.099 & 0.086\\
    EIN\cite{ein_v2} & 26.0M&1,2&0.134&0.088 &  0.187 & 0.183 & 0.256 & 0.175 & 0.186 & 0.257 \\
    \hline
    SoundDet\_backbone(foa) &  10.0 M&0&0.040&0.025 &  0.055 & 0.090 & 0.096 & 0.066 & 0.051 & 0.033 \\
    SoundDet\_backbone(mic) &  10.0 M&0&0.035&0.021 &  0.091 & 0.090 & 0.094 & 0.065 & 0.049 & 0.027 \\
    SoundDet\_nomaxcorr(foa) &  10.0 M&0&0.025&0.015 &  0.034 & 0.080 & 0.063 & 0.040 & 0.032 & 0.014 \\
    SoundDet\_nomaxcorr(mic) &  10.0 M&0&0.017&0.059 &  0.016 & 0.044 & 0.041 & 0.017 & 0.079 & 0.016 \\
    \hline
    SoundDet\_nomots(foa) &  13.0 M&0&0.117&0.062 &  0.173 & 0.152 & 0.247 & 0.174 & 0.159 & 0.232 \\
    SoundDet(foa) &  13.0 M&0&\textbf{0.197}&\textbf{0.098} & \textbf{0.201} & \textbf{0.216} & \textbf{0.294} & \textbf{0.189} & \textbf{0.223} & \textbf{0.289} \\
    \hline
    \end{tabular}
    \label{table_event_based_eval}
\end{table*}

The segment-based evaluation result is shown in Table\,\ref{table_segment_based_eval_new}. We can see that, by involving MaxCorr filter banks, our backbone-only SoundDet achieves comparable performance comparing with SELDNet\cite{seld_dcase19}. SELDNet extracts Log-mel, GCC-PHAT and Intensity features\,(see Table\,\ref{table_event_based_eval}) as its neural network input, replacing these hand-crafted features with learnable MaxCorr filter bank helps to achieve similar performance, even with 1D convolution. If disabling SoundDet to learn between-channel correlation\,(SoundDet\_nomaxcorr), we can witness a huge performance drop in both FOA and MIC format, especially the high event detection error rate and large DoA error angle. This shows that our proposed MaxCorr filter bank is capable of learning essential representations for recovering temporal and spatial information. Actually, we observed a fast convergence of the between-channel time shift parameters\,(see Eqn.\ref{maxcorr-filter}), SoundDet quickly updates time shift parameters in the very first couple of training iterations and swiftly achieves a relatively stable state. Two learned MaxCorr filters are shown in Fig.\ref{fig:learned_maxcorr_filter}, from which we can observe that different MaxCorr filters have learned different time shifts and frequency cutoffs. These learned frequency-selective and phase-sensitive MaxCorr filter bank is of vital importance for sound event representation learning.

Both SoundDet and SELDNet work better for FOA format than MIC format. Framewise based learning\,(both SoundDet, SELDNet and EIN) shows no obvious difference in estimating events with different temporal length, which is reasonable because of their framewise processing property. The whole SoundDet\,(combine backbone and two heads) far outperforms SELDNet baseline and achieves comparable performance with EIN with much less parameter number. Specifically, SoundDet shows advantage on longer sound event estimation. It thus attests the necessity of directly modelling ``sound-object" when it comes to events with longer temporal length or more complex motion trajectory. Excluding motion smoothness largely reduces the performance, we observed involving motion smoothness map greatly helps the neural network to accurate localize sound events.

The event-based evaluation result is shown in Table\,\ref{table_event_based_eval}, from which we can observe that SoundDet outperforms all existing methods in both the mAP and mAR metric. This shows that the segment-based metrics in current use do not comprehensively reflect an algorithm's performance. Rather than focusing on a local segment evaluation, the event-based metric highlights an event's completeness and continuity and SoundDet is naturally designed to meet these requirements. This advantage is echoed by the qualitative comparison in Fig.\,\ref{qualitative_result}. We can clearly see that SELDNet and EIN inevitably contain many mixed sound events and frequently cut complete events into small disconnected segments. The situation becomes more serious when faced with overlapping situation. However, SoundDet better avoids this dilemma because by design it treats events as complete instances, rather than discrete segments. 

\begin{table}[h]
    \centering
    \small
    \caption{Inference time on Intel(R) Core(TM) i9-7920X CPU. The waveform pre-processing time is contained for SELDNet and EIN.}
    \begin{tabular}{|c|c|c|}
        \hline
        SELDNet & EIN & SoundDet \\ 
        \hline
        1.20~s & 2.20~s & 1.25~s \\
        \hline
    \end{tabular}
    
    \label{inference_time_test}
\end{table}

We further report the average inference time of different methods to process a one-minute long audio in Table.\,\ref{inference_time_test}, showing that it is almost twice as fast as EIN, and comparable to SELDNet. In summary, SoundDet is capable of learning a sound event's spatial, temporal and class information from raw waveforms, achieving competing performance under existing segment-based metrics and leading results on the proposed event-based metrics.

\section{Conclusion}
We have introduced a number of innovations in this paper that attempt to unify the currently disparate fields of object detection in computer vision, with event detection and localization in audio settings. We also abandon prior approaches that resort to hand-crafted pre-processing and transformation of the multi-channel waveforms, and instead directly consume raw data. To this end, we propose a more object-centric approach to measuring the performance of sound event detection and localization through metrics of mAP and mAR. It is our hope that this alternative view will seed further development in this space. One potential future research direction is to design more elegant learnable filter banks.

\section{Acknowledgement}
We thank Qi Ou from Department of Earth Science, University of Oxford for helpful discussion. We also appreciate reviewers' constructive feedback to improve the work. 

\begin{table}[H]
    \centering
    \small
    \caption{SoundDet architecture illustration. The layer follow $name@kernelsize,stride$ format. MaxCorr here indicates MaxCorrelation filter bank. cn indicates class number. FC is fully connection layer. B is the batchsize, T is the waveform temporal length. All convolution layers are followed by a batch normalization layer and Relu activation layer and FC layers are followed by leaky Relu activation layer.}
    \begin{tabular}{|c|c|c|}
        \hline
        layer & filter num  & output size \\ 
        \hline
        \multicolumn{3}{|c|}{Input: [B,4,T]}\\
        \hline
        \multicolumn{3}{|c|}{SoundDet Backbone Network} \\
        \hline
        MaxCorr@251,75 & 256 & [B, T/75,256]\\
        \hline
        conv1d@3,2 & 128 & [B, T/150,128] \\
        \hline
        conv1d@3,2 & 128 & [B, T/300,128] \\
        \hline
        conv1d@3,2 & 256 & [B, T/600,256] \\
        \hline
        conv1d@3,2 & 256 & [B, T/1200,512] \\
        \hline
        conv1d@3,2 & 512 & [B, T/2400,512] \\
        \hline
        conv1d@3,2 & 512 & [B, T/4800,512] \\
        \hline
        conv1d@3,2 & 1024 & [B, T/9600,1024] \\
        \hline
        deconv1d@3,2 & 512 & [B, T/4800,512] \\
        \hline
        deconv1d@3,2 & 512 & [B, T/2400, 512] \\
        \hline
        biGRU & 512 & [B, T/2400,512] \\
        \hline
        conv1d@3,1 & 512 & [B, T/2400,512] \\
        \hline
        \multicolumn{3}{|c|}{Event Multilabel Classify} \\
        \hline
        FC & 512 & [B, H, W, 512] \\
        \hline
        FC & 256 & [B, H, W, 256] \\
        \hline
        FC & class num & [B, H, W, cn] \\
        \hline
        \multicolumn{3}{|c|}{Eventness Classify} \\
        \hline
        FC & 256 & [B, H, W, 256] \\
        \hline
        FC & 1 & [B, H, W, 1] \\
        \hline
        \multicolumn{3}{|c|}{tIoU Map Regression} \\
        \hline
        FC & 256 & [B, H, W, 256] \\
        \hline
        FC & 1 & [B, H, W, 1] \\
        \hline
        \multicolumn{3}{|c|}{Spatial Location Head} \\
        \hline
        FC & 512 & [B, T/2400, 512] \\
        \hline
        FC & 512 & [B, T/2400, 512] \\
        \hline
        FC & 256 & [B, T/2400, 256] \\
        \hline
        FC & classnum$\times$3 & [B,T/2400,cn$\times$3] \\
        \hline
    \end{tabular}
    \label{SoundDet_network}
\end{table}

\bibliography{main}
\bibliographystyle{icml2021}





\end{document}